\documentclass[graybox, natbib, nosecnum, twocolumn]{svmult}
\bibpunct{(}{)}{;}{a}{}{,} 

\usepackage{mathptmx}       
\usepackage{helvet}         
\usepackage{courier}        
\usepackage{type1cm}        

\usepackage{makeidx}         
\usepackage{graphicx}        
\usepackage{multicol}        
\usepackage[bottom]{footmisc}
\usepackage[normalem]{ulem}	
\usepackage{hyperref}  
\usepackage{breakurl}
\usepackage{soul}   

\usepackage{multirow}
\usepackage{booktabs}
\usepackage{enumitem}
\usepackage{comment}

\newcommand{\repeatthanks}{\textsuperscript{\thefootnote}}

\begin{document}

\title*{Source Address Validation}
\titlerunning{SAV}

\author{Maciej Korczy\'nski \thanks{Univ. Grenoble Alpes, CNRS, Grenoble INP, LIG, France} and Yevheniya Nosyk\repeatthanks}

\authorrunning{Korczy\'nski and Nosyk}

\institute{Maciej Korczy\'nski \at Univ. Grenoble Alpes, CNRS, Grenoble INP, LIG, France, \email{maciej.korczynski@univ-grenoble-alpes.fr} \and Yevheniya Nosyk \at Univ. Grenoble Alpes, CNRS, Grenoble INP, LIG, France \email{yevheniya.nosyk@etu.univ-grenoble-alpes.fr}}


%
\maketitle
%

\section{Definitions}
Source Address Validation (SAV) is a standard formalized in RFC 2827 aimed at discarding packets with spoofed source IP addresses.
The  absence  of  SAV  
has  been  known  as  a root  cause  of  reflection Distributed  Denial-of-Service  (DDoS)  attacks.

\textit{Outbound SAV (oSAV)}: filtering applied at the network edge to traffic coming from inside the customer network to the outside.

\textit{Inbound SAV (iSAV)}: filtering applied at the network edge to traffic coming from the outside to the customer network.

\section{Background }
The Internet relies on IP packets to enable communication between hosts with the destination and source addresses specified in packet headers. However, there is no packet-level authentication mechanism to ensure that the source address has not been altered~\citep{Beverly:2009:UED:1644893.1644936}. The modification of a source IP address is referred to as ``IP spoofing''. It results in the anonymity of the sender and prevents a packet from being traced to its origin. This vulnerability has been leveraged to launch Distributed Denial-of-Service (DDoS) attacks that can be made even more effective using reflection~\citep{bb-spoofer-sruti}. Because it is not possible in general to prevent packet header modification, concerted efforts have been undertaken to prevent spoofed packets from reaching potential victims. This goal can be achieved by filtering packets at the network edge, formalized in RFC~2827, and called \textit{Source Address Validation} (SAV)~\citep{Ferguson:2000:NIF:RFC2827}.

The RFC defined the notion of ingress filtering---discarding any packets with source addresses not following filtering rules. This operation is the most effective when applied at the network edge~\citep{Ferguson:2000:NIF:RFC2827}. RFC 3704 proposed different ways to implement SAV including static access control lists (ACLs) and reverse path forwarding \citep{Baker:2004:IFM:RFC3704}. 

Packet filtering can be applied in two directions: \textit{inbound} to the customer's network from outside \citep{korczyski2020dont} and \textit{outbound} from the customer to outside \citep{Ferguson:2000:NIF:RFC2827}.
The lack of SAV in any of these directions may result in different~security~threats.

Attackers benefit from the absence of 
oSAV to launch DDoS attacks, in particular, reflection attacks. Adversaries make use of public services prone to amplification~\citep{hell}, such as open DNS resolvers or NTP servers, to which they send requests on behalf of their victims by spoofing their source IP addresses. The victim is then overloaded with the traffic coming from the services rather than from the botnet controlled by the attacker.
In this scenario, the origin of the attack is not traceable. One of the most successful attacks against GitHub resulted in traffic of 1.35 Tbps: attackers redirected Memcached responses by spoofing their source addresses~\citep{github}. In such scenarios, spoofed source addresses of the victims are usually globally routable IPs. In some cases, to impersonate an internal host, a spoofed IP address may be from the inside target network, which reveals the absence of iSAV~\citep{Baker:2004:IFM:RFC3704}.

Pretending to be an internal host reveals information about the inner network structure, such as the presence of closed DNS resolvers that resolve only on behalf of clients within the same network.
The absence of iSAV 
may have serious consequences when combined with the NXDOMAIN attack, also known as the Water Torture Attack~\citep{LuoWXCYT18}, or the recently discovered NXNSAttack~\citep{NXNSAttack}.
Both attacks enable Denial-of-Service against both recursive resolvers and authoritative servers.

The possibility of impersonating another host on the victim network can also assist in the zone poisoning attack~\citep{Korczynski:2016:ZPN:2987443.2987477}. A 
DNS server, authoritative for a given domain, may be configured to accept so-called non-secure DNS dynamic updates from 
hosts (e.g. a DHCP server) on the same network \citep{rfc2136}. Therefore, sending a single spoofed 
UDP packet from the outside with an IP address of that 
host will modify the content of the zone file \citep{Korczynski:2016:ZPN:2987443.2987477}. 
The attack vector can be used to hijack the domain name. 
Another way to target closed resolvers is to perform DNS cache poisoning~\citep{kaminsky}. An attacker can send a spoofed DNS \texttt{A} request for a specific domain to a closed resolver, followed by forged replies before the arrival of the response from the genuine authoritative server. In this case, the users who query the same domain will be redirected to where the attacker specified until the forged DNS entry reaches~its Time To Live~(TTL).

Despite the knowledge of the above-mentioned attack scenarios and the costs of the damage they may incur, it was shown that SAV is not yet widely deployed. Lichtblau et al surveyed 84 network operators to learn whether they deployed SAV and what challenges they faced~\citep{Lichtblau}. The reasons for not performing packet filtering included incidentally filtering out legitimate traffic, equipment limitations, and lack of a direct economic benefit. In the case of outbound SAV, the compliant network cannot become an attack source but can be attacked itself. Therefore, oSAV suffers from misaligned economic incentives: a network operator that adopts oSAV incurs the cost of deployment, while the security profits 
benefit all other networks \citep{saving}.
On the other hand, performing inbound SAV protects networks from direct threats, which is beneficial from an economic perspective.

\section{Application}

Given the prevalent role of IP spoofing in cyberattacks, there is a need to estimate the level of SAV deployment by network providers.  
Increasing the visibility of the networks that allow spoofing leads to a decrease in the information asymmetry between network operators, their peers and customers and thus may strengthen the economic incentives for the adoption of SAV.

\begin{table*}[t]
\caption{Methods to infer  deployment of SAV} 
\label{related_work} 
\scriptsize
\centering 
\setlength{\tabcolsep}{6pt}
\begin{tabular}{lcccc}
 \toprule
\multirow{3}{5em}{\textbf{Method}} &
\multirow{3}{4em}{\textbf{SAV direction}} &
\multirow{3}{4em}{\textbf{Presence/ absence}} &
\multirow{3}{3em}{\textbf{Remote}} &
\multirow{3}{5em}{\textbf{Relies on misconfigurations}} \\
&&&&\\
&&&&\\
 \midrule
  Closed Resolver~\citep{korczyski2020dont}& iSAV & both & yes & no\\
 Spoofer~\citep{bb-spoofer-sruti} & oSAV/iSAV & both & no & no\\ 
 Forwarder-based~\citep{Kuhrer:2014:EHR:2671225.2671233} & oSAV & absence & yes & yes\\
 Traceroute loops~\citep{loops} & oSAV & absence & yes & yes\\
 Spoofer-IX~\citep{Lucas} & oSAV & both & no & no \\
 \bottomrule
\end{tabular}
\end{table*}

Table~\ref{related_work} summarizes methods proposed to infer SAV deployment. They differ in terms of the filtering direction 
(iSAV versus oSAV) whether they infer the presence or absence of SAV, whether measurements can be done remotely or on a vantage point inside the tested network is required, and if the method relies on existing network misconfigurations.

The Closed Resolver project \citep{10.1145/3404868.3406668,korczyski2020dont,DBLP:journals/corr/abs-2006-05277} aims at mitigating the problem of inbound IP spoofing. 
They identify closed and open DNS resolvers that accept spoofed requests coming from the outside of their network.
The proposed method is remote and does not rely on existing misconfigurations.
It provides the most complete picture of iSAV deployment by network providers and covers over 55 \% IPv4 and 27 \% IPv6 ASes. It reveals that the great majority of ASes are fully or partially vulnerable to inbound spoofing.

The Spoofer project \citep{bb-spoofer-sruti,Beverly:2009:UED:1644893.1644936,spoofer_new} deploys a client-server infrastructure mainly based on volunteers and ``crowdworkers'' hired for one study trough five crowdsourcing platforms~\citep{marketplaces} that run the client software from inside a network. The active probing client sends both unspoofed and spoofed packets to the Spoofer server either periodically or when it detects a new network. The server inspects received packets (if any) and analyzes whether spoofing is allowed and to what extent~\citep{Beverly:2009:UED:1644893.1644936}. 
This approach identifies the absence and the presence of SAV in both directions. The results obtained by the Spoofer project provide the most confident picture of the deployment of oSAV and have covered  tests  from 7,915 ASes since 2015 \citep{Spoofer}. 
However, those that are not aware of this issue or do not deploy oSAV are less likely to run Spoofer on their networks.

A more practical approach is to perform such measurements remotely. \cite{Kuhrer:2014:EHR:2671225.2671233} scanned for open DNS resolvers, as proposed by~\cite{mauch}, to detect the absence of outbound SAV. The method leverages the misconfiguration of forwarding resolvers and is referred to as \textit{forwarder-based}.
The misbehaving resolver forwards a request to a recursive resolver with either not changing the packet source address to its own address or by sending back the response to the client with the source IP of the recursive resolver. 
Misconfigured forwarders revealed 2,692 ASes that are fully or partially vulnerable to outbound spoofing.

\cite{loops} proposed another method that does not require a vantage point inside a tested network. When packets are sent to a customer network with an address that is routable but not allocated, this packet is sent back to the provider router without changing its source IP address. The packet, having the source IP address of the machine that sent it, should be dropped by the router because the source IP does not belong to the customer network. 
The method detected 703 ASes not deploying oSAV.

Finally, while the above-mentioned methods rely on actively generated (whether spoofed or not) packets, \cite{Lucas} passively observed and analyzed inter-domain traffic exchanged between 
networks at a large IXP 
taking into account AS business relationships, asymmetric routing, and traffic engineering.

\section{Open Problems and Future Directions}
Although the Internet community has developed technical solutions to mitigate the spoofing vulnerability and a variety of methods to estimate the level of SAV deployment by network providers, its deployment remains low.
Lack of a direct economic benefit in case of deploying oSAV remains one of the primary problems preventing providers from applying the existing technical standards~\citep{Lichtblau}.
This failure is referred to as \textit{negative externality}: network operators do not invest in implementing the security standard while imposing economic costs on other networks that are victims of attacks using IP spoofing~\citep{spoofer_new}.

The deployment of 
iSAV does not suffer from misaligned economic incentives and protects  the  provider  network that deploys the standard rather  than other networks.
Interestingly, SAV for outbound traffic turned out to be more deployed than inbound at the AS level among network operators committed to the Mutually Agreed Norms for Routing Security
regulations~\citep{manrs} initiative \citep{spoofer_new,DBLP:journals/corr/abs-2006-05277}. 
At the time of writing, 515 ASes are its signatories. MANRS requires its members to implement SAV in their networks ``to prevent packets with an incorrect source IP address from entering or leaving the network"~\citep{DBLP:journals/corr/abs-2006-05277}.
One possible explanation for the higher deployment of oSAV among MANRS members is that the absence of outbound packet filtering gained widespread attention since it is the reason for reflection DDoS attacks. Under these circumstances, the SAV of inbound traffic remained neglected or overlooked by network operators.

``Naming and shaming'' of network operators appeared to be a weak form of incentive~\citep{spoofer_new} for deploying oSAV. \cite{spoofer_new} consider several potential future scenarios, including liability associated
with attacks originating from their networks or different types of regulations, including governmental initiatives.
Finally, long-term efforts taken by the research community to measure and notify non-compliant operators such as the Spoofer project for oSAV and the Closed Resolver project for iSAV may significantly contribute to improving the overall deployment of the standard.  

\bibliographystyle{spbasic}  
\bibliography{author} 

\end{document}